\documentstyle[pra,aps,graphicx]{revtex}
\begin{document}
\draft
\title{Anharmonicity  of  Flux Lattices  and Thermal Fluctuations in
Layered Superconductors} 
\author{T.Ruth Goldin and Baruch Horovitz} 
\address{Department of Physics, Ben-Gurion University,  Beer-Sheva 
84105, Israel}
\maketitle
\begin{abstract}
We study elasticity of a perpendicular flux lattice in a layered 
superconductor with Josephson coupling between layers. We  find 
that  for flux displacements $\rho$ the energy contains 
$\rho^{2}\ln\rho$ terms, so that elastic constants cannot be 
strictly defined. Instead we define effective elastic
constants by a thermal average. The tilt moduli have terms $\sim 
\ln T$ which for $\lambda_{J}\ll a$, where $\lambda_{J}$ is the 
Josephson length and $a$ is the flux line spacing, lead to $\langle 
\rho ^{2} \rangle \sim T/|\ln T|$. The expansion parameter indicates 
that the dominant low temperature phase transition is either layer 
decoupling at high fields ($\lambda_{J}\gg a$) or melting at low 
fields ($\lambda_{J}\ll a$).
\end{abstract}

\section{Introduction}
 
The properties of Abrikosov flux lattices in layered superconductors 
are of considerable interest in view of numerous
experiments on high $T_{c}$ compounds \cite{Kes}. For magnetic field 
perpendicular to the layers, the flux lattice can be
considered as two dimensional point vortices in each superconducting 
layer which are stacked one on top of the other. Each point vortex, or 
a pancake vortex, represents a singularity of the superconducting 
order parameter, i.e. the superconductor's phase in a given layer
changes by $2\pi$ around the vortex. These pancake vortices are 
coupled by their magnetic field as well as by the 
Josephson tunneling between nearest layers. The
fluctuations of the displacements of pancake vortices are manifested by a 
variety of experiments \cite{Blatter},  and 
affect phase transitions such as melting of the flux lattice and layer 
decoupling, i.e. vanishing of the interlayer Josephson coupling on long 
scales \cite{Glazman,Daemen,Baruch}.
  
A harmonic expansion of the flux lattice energy to second order in 
displacements of the flux lines, defines in
general the elastic constants \cite{Brandt}. For anisotropic layer 
systems a harmonic expansion was studied by Glazman and
Koshelev (GK) \cite{GK}. This expansion is, however, nontrivial, 
since it involves expanding the nonlinear Josephson
coupling $J\cos \phi_{n,n+1}$ where $\phi_{n,n+1}$ is the relative 
phase of neighboring layers. When two pancake vortices
which are one on the top of the other are separated by a distance 
$2\rho$ (parallel to the layers) then $\phi_{n,n+1}$ has
large variations in a circle of radius $\rho$ between these pancake
vortices. This effect has led GK \cite{GK} to anticipate a 
$\rho^{2}\ln\rho$ term in the energy expansion when $\rho > \xi$ 
where $\xi$ is the in-layer coherence length. This term was also 
found by Kramer \cite{Kramer} for a single vortex line and 
independently by us \cite{Goldin}.

In this work we present the detailed energy expansion. In 
addition to the difficulty at short scales, leading to the
$\rho^{2}\ln\rho$ term, we find that the convergence parameter of the 
expansion vanishes as $1/|\ln E_J|$ when the Josephson
coupling $E_J\rightarrow 0$. Thus the $E_J=0$ elastic constants are not 
recovered by a $E_J\rightarrow 0$ limit. 
We then show (Section III D) how to define effective elastic 
constants and apply our results (Section IV) to
thermal averages of $\langle\rho^{2}\rangle$. In particular we find 
$\langle \rho ^{2} \rangle \sim T/|\ln T|$ for $\lambda_{J}\ll a$,
 where $\lambda_{J}$ is the Josephson length and $a$ is the flux 
line spacing. The expansion parameter indicates the type of dominant 
fluctuation with a related instability, i.e. for $\lambda_{J}\gg a$
the decoupling transition dominates while for $\lambda_{J}\ll a$ the 
melting transition dominates. The relevance to experimental data is 
discussed in section V.

\section{The Model}
We start from the  Lawrence-Doniach \cite{Lawrence} free energy in 
terms of superconducting phases $\phi_{n}({\bf r})$ on the
$n$-th layer and the vector potential $\vec{A}({\bf r},z)$:
\begin{eqnarray}
{\cal F}&=&\frac{1}{8\pi}\int d^{2}{\bf r} d z
 \left[
(\vec{\nabla}\times\vec{A}({\bf r},z)          )^{2}+
\frac{d}{\lambda_{a b}^{2}}\sum_{n}
(\frac{\Phi_{0}}{2\pi}{\bf\nabla}\phi_{n}({\bf r})-\vec{A}({\bf r},z)          
)^{2} \delta(z-n d) \right]    \nonumber \\
              &-&E_J\sum_{n}\int d^{2}{\bf r}
\cos\left[          \phi_{n}({\bf r})-\phi_{n-1}({\bf r})-
\frac{2\pi}{\Phi_{0}} \int_{(n-1) d}^{nd} A_{z}({\bf r},z) d z           
\right]                                                  \label{LD}
\end{eqnarray}
where  $\lambda_{ab}$ is the penetration length parallel to the 
layers, $d$ is the spacing between 
layers and $\Phi_{0}=hc/2e$ is the flux quantum.

In Appendix A we consider a simplified model with two layers in the 
$e\rightarrow 0$
(i.e. $\vec{A}({\bf r},z)$ decouples from $\phi_{n}({\bf r})$) limit and two 
pancake vortices, one at the bottom
layer at ${\bf R}_{1}$, and the other at the top layer at 
${\bf R}_{1}+{\mbox{\boldmath $\rho$}}$. It is rather straightforward, as 
shown in
the Appendix A, to show that the free energy expansion has a 
$\rho^{2}\ln \rho$ term.

We proceed with the  partition sum for Eq.\ (\ref{LD}) which involves 
integrating over $\phi_{n}({\bf r})$ and $\vec{A}({\bf r},z)$,  subject to a 
gauge condition.
 Since $\vec{A}({\bf r},z)$ is a Gaussian field (choosing the axial gauge 
$A_{z}({\bf r},z)=0$, $\vec{A}({\bf r},z)=({\bf A}({\bf r},z),0)$ ) we can shift 
 ${\bf A}\rightarrow {\bf A}+\delta {\bf A}$ where ${\bf A}({\bf r},z)$ 
now satisfies the $x$, $y$ components of
\begin{equation}
\vec{\nabla}\times\vec{\nabla}\times {\bf 
A}({\bf r},z)=\frac{d}{\lambda_{a b}^{2}}\sum_{n}
[       \frac{\Phi_{0}}{2\pi}{\bf\nabla}\phi_{n}({\bf r})-{\bf 
A}({\bf r},z)          ]\delta(z-n d)                 \label{Maxwell}
\end{equation} 
and then fluctuations in $\delta {\bf A}$ decouple from  those of 
$\phi_{n}({\bf r})$. The partition sum at temperature $T$ is now 
\begin{equation}
{\cal Z}= \int {\cal D}\phi_{n}({\bf r}) \exp [-{\cal F}/T]
\end{equation}
with $\vec{A}({\bf r},z)$ in Eq.\ (\ref{LD}) given by the solution of 
Eq.\ (\ref{Maxwell}). Note 
that since Eq.\ (\ref{Maxwell}) is gauge invariant under 
${\bf A}\rightarrow {\bf 
A}-\frac{\Phi_{0}}{2\pi}{\bf\nabla}\chi({\bf r},n d)$ and
 $\phi_{n}({\bf r}) \rightarrow  \phi_{n}({\bf r})-\chi({\bf r},n d)$
 one can in  fact  choose any gauge.

We now decompose  $ \phi_{n}({\bf r})$ to
\begin{mathletters}
\begin{eqnarray}
 \phi_{n}({\bf r}) &=& \phi_{n}^{0}({\bf r})+
\sum_{{\bf r} '}s_{n}({\bf r} ')\alpha({\bf r}-{\bf r} ')  
\label{decompose}\\
\theta_{n}({\bf r})&=& \phi_{n}^{0}({\bf r})- \phi_{n-1}^{0}({\bf r})
\end{eqnarray}
\end{mathletters}
where $\phi_{n}^{0}({\bf r})$ is the nonsingular part of $\phi_{n}({\bf r})$, 
$\alpha({\bf r})=\arctan (y/x)$, 
$s_{n}({\bf r})=1$ at pancake vortex sites and
$s_{n}({\bf r})=0$ otherwise. The sum in Eq.\ (\ref{decompose}) is 
then a sum on ${\bf r}'$ being the vortex positions on the n-th layer.

Solving Eq.\ (\ref{Maxwell}) for ${\bf A}$ in terms of  
$\theta_{n}({\bf r})$ and 
$s_{n}$, substituting in Eq.\ (\ref{LD}) yields \cite{Horovitz}
\[{\cal F}={\cal F}_{v}+{\cal F}_{J}+{\cal F}_{f} \]
where  ${\cal F}_{v}$ is the vortex-vortex interaction via 
the 3D magnetic field,  ${\cal F}_{J}$ is interlayer Josephson 
coupling term and ${\cal F}_{f}$ is an energy due to fluctuations of 
the nonsingular  phase:
\begin{mathletters}
\begin{eqnarray}
  {\cal F}_{v}&=&\frac{1}{2}\sum_{r,n} \sum_{r',n'}s_{n}({\bf r})
G_{v}({\bf r}-{\bf r} ';n-n')s_{n'}({\bf r} ') \\
{\cal F}_{J}&=&-E_J\sum_{n} \int d^{2}{\bf r}
(\cos[\theta^{n}({\bf r}) + \sum_{{\bf r} '}(s_{n}({\bf r} ')-s_{n-1}({\bf r} '))
\alpha({\bf r}-{\bf r} ')]-1)                             \label{FJos}\\
    {\cal F}_{f}&=&\frac{1}{2}\sum_{q,k} G_{f}^{-1}(q,k) 
|\theta({\bf q},k)|^{2}                            \label{Ff} 
\end{eqnarray}  
\end{mathletters}
Here $({\bf q},k)$ is a 3D wave vector,  
$\hat{z}$ is the direction perpendicular to the plane of layers and
\begin{mathletters}
\begin{eqnarray}
G_{v}(q,k)&=&\frac{\Phi_{0}^{2}d^{2}}{4\pi\lambda_{a b}^{2}} 
\frac{1}{q^{2}}\frac{1}{1+f(q,k)}          \\
    f(q,k)&=&\frac{d}{4\lambda_{a b}^{2}q}\frac{\sinh qd}
{\sinh^{2}\frac{qd}{2}+\sin^{2}\frac{kd}{2}}   \\
G_{f}(q,k)&=&\frac{16\pi^{3}d^{2}}{\Phi_{0}^{2}q^{2}}\left(1+
\frac{4\lambda_{a b}^{2}} {d^{2}} \sin^{2}\frac{kd}{2}\right) 
\end{eqnarray}
\end{mathletters}

For deviations ${\bf u}_{l}^{n}$ of 2D vortices on the
$n$-th layer
from equilibrium positions ${\bf R}_{l}$ of a hexagonal lattice , the
function $s_{n}({\bf r})$ is
\[s_{n}({\bf r})=\left\{ \begin{array}{ll}
1 & \mbox{if ${\bf r}={\bf R}_{l}+{\bf u}_{l}^{n}$}\\
0 & \mbox{otherwise}\, . \end{array} \right. \]

The Fourier transform
\[{\bf u}({\bf q},k)=\sum_{n,l}{\bf u}_{l}^{n} \exp (i {\bf q} {\bf 
R}_{l}+i k n d)\]
identifies longitudinal $u^{l}({\bf q},k)={\bf q} \cdot {\bf 
u}({\bf q},k)/q$ and transverse $u^{tr}({\bf q},k)=[{\bf q} 
\times\hat{z}] \cdot {\bf u}({\bf q},k)/q$ components of ${\bf 
u}({\bf q},k)$
\[{\bf u}({\bf q},k)=u^{l}({\bf q},k){\bf q}/q+u^{tr}({\bf q},k)[{\bf 
q} \times\hat{z}]/q  \, .\]  

If the free energy can be expanded to second order in ${\bf u}({\bf 
q},k)$ then 
the compression $c_{11}$, shear $c_{66}$ and tilt
$c_{44}$ moduli for the vortex lattice are identified by: 
               
\begin{eqnarray}
{\cal F}
   &=&\frac{1}{2}\int_{BZ} \int^{\pi /d}\frac{d^{2} {\bf q} 
dk}{(2\pi)^{3}} (d a^{2})^{2}
   \left\{[q^{2} c_{11}(q,k)+k_{z}^{2} c_{44}^{l}(q,k) ]
   |u^{l}({\bf q},k)|^{2} + \right.  \nonumber\\
   & &\left. [q^{2} c_{66}(q,k) + k_{z}^{2} c_{44}^{tr}(q,k)]
   |u^{tr}({\bf q},k)|^{2}
   \right\}
\end{eqnarray}
where $ k_{z}^{2}=\frac{4}{d^{2}}\sin^{2}\frac{kd}{2} $, $a^{2}$ is 
the area of a unit cell ($a^{2}=\Phi_{0}/B$) and
$\int_{BZ}$ is for ${\bf q}$ integration over the Brillouin zone.
We assume below that $d\ll a, \lambda_{ab}$ as is the case for high 
$T_{c}$ compounds \cite{Kes,Blatter}.

Note that for ${\bf q} = 0$ there should be no distinction between 
transverse and longitudinal 
$c_{44}^{l}(0,k)=c_{44}^{tr}(0,k)$, however for $q\neq 0$, 
$c_{44}^{l}(q,k)$ and $c_{44}^{tr}(q,k)$ may differ.

\section{Elastic Constants of the Flux Lattice}

\subsection{Magnetic Coupling}

We consider first the case with no Josephson coupling, $E_J=0$.
The vortex-vortex interaction has then  the form

\begin{eqnarray*}
{\cal F}_{v}&=&\frac{1}{2} \sum_{n,l} \sum_{n',l'}(1-\delta_{l,l'})
\int^{\infty} \int_{-\pi/d}^{\pi/d}
\frac{d^{2} {\bf q} dk}{(2\pi)^{3}} G_{v}(q,k) e^{i{\bf q}({\bf 
R}_{l}-{\bf R}_{l'}+{\bf u}_{l}^{n}-{\bf u}_{l'}^{n'})} e^{ik(n-n')d} 
+ \\
       & &+\frac{1}{2}\sum_{l}
\sum_{n,n'}(1-\delta_{n,n'}) \int^{\infty} \int_{-\pi/d}^{\pi/d}
\frac{d^{2} {\bf q} dk}{(2\pi)^{3}} G_{v}(q,k) e^{i{\bf q}({\bf 
u}_{l}^{n}-{\bf u}_{l}^{n'})} e^{ik(n-n')d}
\end{eqnarray*}  

The first and the second terms can be expanded with respect to   
${\bf q}\cdot {\bf u}$ since in absence of the $l=l'$ , $n=n'$
term they converge. It is
important {\em not} to decompose the $(1-\delta_{l,l'})$ or      
$(1-\delta_{n,n'})$
factors until all integrals converge; the $l=l'$, $n=n'$ terms 
produce then the integral terms in the following elastic matrix
 
\begin{eqnarray*}
         {\cal F}_{v}&=&\frac{1}{2}\int_{BZ}\int^{\pi /d} \frac{d^{2} 
{\bf q} dk}{(2\pi)^{3}}
(d a^{2})^{2} \phi^{ij}({\bf q},k)u_{i}({\bf q},k)
u_{j}^{*}({\bf q},k)\\ 
\phi^{ij}({\bf q},k) &=&(\frac{1}{d a^{2}})^{2}  \sum_{{\bf Q}} \left[
G_v (|{\bf Q}-{\bf q}|,k)({\bf Q}-{\bf q})^{i}({\bf Q}-{\bf q})^{j} - 
G_v (Q,0)Q^{i}Q^{j}
\right]  \\
                     &-&(\frac{1}{d 
a^{2}})^{2}\int^{\infty}\frac{d^{2}{\bf p}a^{2}}{(2\pi)^{2}} \left[
                G_v (|{\bf p}-{\bf q}|,k)({\bf p}-{\bf q})^{i}({\bf 
p}-{\bf q})^{j} -   
G_v (p,0)p^{i}p^{j} \right] \\
                     &+&\frac{\delta_{ij}}{2} (\frac{1}{d a^{2}})^{2} 
\int^{\infty}
\frac{d^{2}{\bf p}a^{2}}{(2\pi)^{2}} p^{2}
\{G_v (p,k)-G_v (p,0) \}
\end{eqnarray*}
where ${\bf Q}$ are 2D reciprocal vectors of the hexagonal lattice.

Considering ${\bf q} \rightarrow 0$ we 
use the symmetry of the hexagonal lattice
\begin{mathletters}
\label{hex}
\begin{eqnarray}
\sum_{i,j}     g(Q)Q_{i}Q_{j}          &=& \frac{1}{2} \sum_{{\bf Q}} 
g(Q)Q^{2}  \\
\sum_{i,j,l}   g(Q)Q_{i}Q_{j}Q_{l}     &=&0 \\
\sum_{i,j,l,m} g(Q)Q_{i}Q_{j}Q_{l}Q_{m}&=&\frac{1}{8}  
(\delta_{ij}\delta_{lm}+\delta_{il}\delta_{jm}+\delta_{im}\delta_{jl})
\sum_{{\bf Q}}g(Q)Q^{4}
\end{eqnarray}
\end{mathletters}
and separate the ${\bf Q}=0$ and $\sum_{{\bf Q}\neq 0}$  parts.
We consider flux line spacing $a\gg d$, so that the ${\bf Q}$ sums 
involve many terms ($\sim (a/d)^{2}$) and the sums can be
approximated by integrals,
\begin{equation}
 \sum_{{\bf Q}\neq 0}\approx\frac{2}{Q_{0}^{2}}\int_{Q_{0}}^{\infty} 
Q dQ                                                 \label{cont}
\end{equation}
where  $Q_{0}^{2}=4\pi B/\Phi_{0}=4 \pi/a^{2}$, i.e. $\pi 
Q_{0}^{2}$ is the area of a Brillouin zone.

Note that for $q \rightarrow 0$ :
\begin{equation}  
G_v (q,k)\rightarrow \frac{\Phi_{0}^{2} d^{2}}{4\pi}\left[ 
\frac{1}{    1+\lambda_{a b}^{2}(q^{2}+k_{z}^{2})     
}+\frac{k_{z}^{2}}{q^{2}}\frac{1}{  
  1+\lambda_{a b}^{2}(q^{2}+k_{z}^{2})     } \right] .
\label{Gv}
\end{equation}
The first term of Eq.\ (\ref{Gv}) contributes to the compression  moduli 
$c_{11}$ while the second term to the longitudinal part of the tilt 
 moduli $c_{44}^{l,0}$. Therefore the $E_J=0$ compression $c_{11}$, 
shear $c_{66}$ and tilt
$c_{44}^{l,0}$,  $c_{44}^{tr,0}$ moduli for $q\ll Q_{0}$ 
are             
\begin{mathletters}   
\begin{eqnarray}
c_{11}(q,k)&=&\frac{B^{2}}{4\pi}\frac{1}{1+\lambda_{a 
b}^{2}(q^{2}+k_{z}^{2})}+ \nonumber \\
           & &\mbox{}+\frac{1}{2}(\frac{1}{d a^{2}})^{2}\left( 
\sum_{{\bf Q}\neq 0}\frac{1}{Q}
              \frac{\partial}{\partial 
Q}[Q^{2}G_v (Q,k)]-\int^{\infty}\frac{a^2 d^2 \!p}{(2\pi)^{2}} 
\frac{1}{p}
               \frac{\partial}{\partial p}[p^{2}G_v (p,k)] \right) 
+\nonumber \\
           & &\mbox{}+\frac{3}{16d^{2}a^{4}}
               \left( \sum_{{\bf Q}\neq 0} \frac{1}{Q}
               \frac{\partial}{\partial 
Q}[Q^{3}\frac{\partial}{\partial Q} G_v (Q,k)]-
               \int^{\infty}\frac{a^2 d^2 \!p}{(2\pi)^{2}} 
\frac{1}{p} \frac{\partial}{\partial p}
               [p^{3}\frac{\partial}{\partial 
p}G_v (p,k)]\right)=\nonumber \\
           &=&\mbox{}\frac{B^{2}}{4\pi}\frac{1}{1+\lambda_{a 
b}^{2}(q^{2}+k_{z}^{2})} -
               \frac{B \Phi_{0}}{(8\pi\lambda_{a b})^{2} }
               \end{eqnarray}

\begin{eqnarray}
c_{66}(q,k)&=&(\frac{1}{d a^{2}})^{2}\frac{1}{16} \left( \sum_{{\bf 
Q}\neq 0} \frac{1}{Q}
\frac{\partial}{\partial Q}[Q^{3}\frac{\partial G_v (q,k)}{\partial 
Q}]-
\int^{\infty}\frac{a^2 d^2 \!p}{(2\pi)^{2}} 
\frac{1}{p}\frac{\partial}{\partial p}[p^{3}\frac{\partial 
G_v (p,k)}{\partial p}] \right)=\nonumber \\
           &=&\mbox{} \frac{B \Phi_{0}}{(8\pi\lambda_{a b})^{2} }\\
 c_{44}^{l,0}(q,k)           
           &=&\mbox{}\frac{B^{2}}{4\pi}\frac{1}{
1+\lambda_{a b}^{2} (q^{2}+k_{z}^{2})
}+c_{44}^{tr,0}(q,k) \label{c440l}\\
           c_{44}^{tr,0}(q,k)
           &=&\frac{1}{2}(\frac{1}{d 
a^{2}})^{2}\frac{1}{k_{z}^{2}}\sum_{{\bf Q}\neq 0} 
[G_v (Q,k)-G_v (Q,0)]Q^{2}=\nonumber \\
           &=&\mbox{}\frac{ 2 B\Phi_{0} }{ (8\pi \lambda_{a b}^{2} 
)^{2} }
               \frac{1}{k_{z}^{2}}\ln\frac{1+k_{z}^{2} 
/Q_{0}^{2}}{1+\xi^{2} k_{z}^{2}}                    \label{c440t}          
\end{eqnarray}
\end{mathletters}

Note that  $c_{44}^{l,0}(q,k)\neq c_{44}^{tr,0}(q,k)$ even for $q 
\rightarrow 0$ due to the singular form of the vortex-vortex 
interaction $G_{v}({\bf q},k)$ (Eq.\ (\ref{Gv})). At $q=0$ the 
$c_{44}$ terms combine into $\frac{1}{2}k_{z}^{2}[ 
c_{44}^{tr,0}(0,k)+c_{44}^{l,0}(0,k)] 
 |{\bf u}(0,k)|^{2}$ which can also be verified by direct expansion 
for ${\bf u}(0,k)$. As shown below, a finite Josephson coupling 
restores the equality $c_{44}^{l}(q,k)=c_{44}^{tr}(q,k)$ at $q 
\rightarrow 0$.

\subsection{Josephson Coupling: "Naive" Expansion}

We consider now the contribution of the Josephson coupling
Eq.\ (\ref{FJos}) to the 
elastic constants by a conventional expansion,
reproducing the results of GK \cite{GK}.
 
The singular part of Josephson phase difference in the interlayer 
Josephson 
coupling term can be written as [see Fig.1]
\begin{equation}
\psi_{l}^{n}({\bf r})=\alpha({\bf r}-{\bf R}_{l}^{n}-
{\mbox{\boldmath $\rho$}}_{l}^{n} )-\alpha({\bf r}-{\bf R}_{l}^{n}+ 
{\mbox{\boldmath $\rho$}}_{l}^{n} ) 
\end{equation}
Here we defined
\begin{mathletters}
\begin{eqnarray} 
                   {\bf R}_{l}^{n}&=&{\bf R}_{l}+\frac{ {\bf 
u}_{l}^{n}+{\bf u}_{l}^{n-1} }{2}\\
{\mbox{\boldmath $\rho$}}_{l}^{n} &=&\frac{{\bf u}_{l}^{n}-{\bf 
u}_{l}^{n-1} }{2} \\
 {\bf v}_{l}^{n}({\bf r})&=&{\bf r}-{\bf R}_{l}^{n} \, .         \label{vln}
\end{eqnarray}
\end{mathletters}

\begin{figure}[htb]
\begin{center}
\includegraphics[bb=20 420 500 700, scale=0.7]{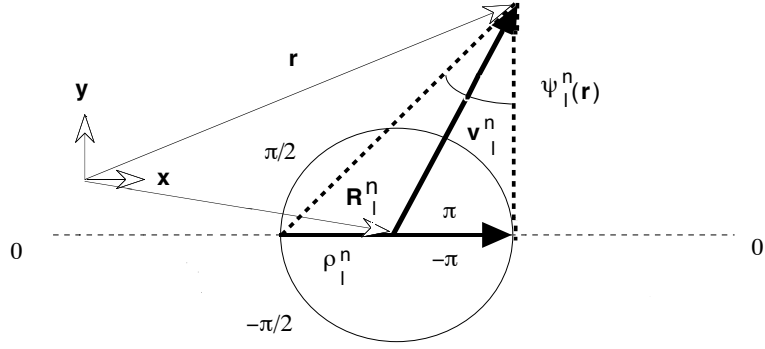}
\end{center}
\caption{The Singular phase difference $\psi_{l}^{n}({\bf r})$ and 
the "$\rho$ circle" where $|{\bf r}-{\bf
R}_{l}^{n}|<\rho_{l}^{n}$ .}
\end{figure}

\vspace{10mm}
The usual way for treatment of the cosine term in Eq.\ (\ref{FJos})
 is by using a "naive" double expansion: 

i) Expansion of the cosine with respect to the phase difference 
\[ \theta^{n}(r)+\sum_{l} \psi_{l}^{n}({\bf r}) \] 
ii) Expansion of the singular phase difference $\psi_{l}^{n}({\bf 
r})$ with respect to 
$\rho_{l}^{n}$
\[ \psi_{l}^{n}({\bf r}) \approx -2{\bf \nabla} \alpha({\bf r}-{\bf 
R}_{l}^{n}) \cdot {\mbox{\boldmath $\rho$}}_{l}^{n}   \]  

Each pair of vortices, displaced by ${\bf u}_{l}^{n}$ and ${\bf 
u}_{l}^{n-1}$ respectively, defines a ``$\rho$-circle'' in space 
where $|{\bf v}_{l}^{n}({\bf r})|<\rho^{n}_{l}$, see Fig. 1. Within a 
``$\rho$-circle''
 $\psi_{l}^{n}({\bf r})$ has a $2\pi$ discontinuity and therefore 
cannot be expanded. The 
expansion (ii) is reasonable only in the region far from the 
``$\rho$-circle'' where $v_{l}^{n}({\bf r}) \gg 
|{\mbox{\boldmath $\rho$}}_{l}^{n}|$. 

Within the approximations (i) and (ii) we can write
\begin{eqnarray}
\lefteqn{\sum_{n} \int d^{2}{\bf r} 
(\cos[\theta^{n}({\bf r}) + \sum_{l}\psi_{l}^{n}({\bf r})]-1)  }
\nonumber\\
& &\approx -\frac{1}{2} \sum_{n} \int d^{2}{\bf r} 
[\theta^{n}({\bf r}) + \sum_{l}\psi_{l}^{n}({\bf r})]^{2}\nonumber \\
& &\approx -\frac{1}{2d} \int \frac{d^{2} {\bf q} dk}{(2\pi)^{3}} ( 
|\theta({\bf q},k)|^{2}+\theta^{*}({\bf q},k)B({\bf q},k)+c.c 
+|B({\bf q},k)|^{2})                                \label{sum}
\end{eqnarray}
where we define
\begin{eqnarray}
          B({\bf q},k) &=&d\sum_{n} \int d^{2}{\bf r} e^{i{\bf q} {\bf r}+ 
iknd}
\sum_{l} B_{l}^{n}({\bf r}) =-\frac{4\pi i d[\hat{z} \times \hat{\bf q}] 
\cdot {\mbox{\boldmath $\rho$}} ({\bf q},k)}{q} 
   \nonumber \\
B_{l}^{n}({\bf r})     &=&-2{\bf \nabla} \alpha({\bf r}-{\bf R}_{l}^{n}) 
\cdot {\mbox{\boldmath $\rho$}}_{l}^{n}=\frac{2 [ 
{\mbox{\boldmath $\rho$}}_{l}^{n}
\times {\bf v}_{l}^{n}({\bf r})]_{z}} {v_{l}^{n}({\bf r})^{2} }   
\end{eqnarray}
and use the Fourier transform
\[ \int d^{2} {\bf r} {\bf \nabla} \alpha({\bf r})e^{i{\bf q} {\bf r}}=
\frac{2\pi i [\hat{z} \times \hat{\bf q}]}{q} \, .\]
Note, that merely the use of the expansion (ii) leads to an error of
order $\rho^{2}$ since the difference 
\( (\psi_{l}^{n}({\bf r}))^{2}-(B_{l}^{n}({\bf r}))^{2}  \) is of order 1 
in the
``$\rho$-circle'' with area $\sim\rho^{2}$.

Combining Eq.\ (\ref{sum}) with ${\cal F}_{f}$ of Eq.\ (\ref{Ff}) yields
\begin{eqnarray}
{\cal F}_{J}+{\cal 
F}_{f}&=&\frac{1}{2}\int^{\infty}\int_{\frac{-\pi}{d}}^{\frac{\pi}{d}}\frac{d^{2} 
{\bf q} dk}{(2\pi)^{3}}(G_{f}^{-1}({\bf q},k)+ E_J/d)
\left|\theta({\bf q},k)-\theta^{0}({\bf q},k)\ \right|^{2}+  
\nonumber  \\
                                 &  &+\frac{E_J}{2d}                  
\int^{\infty}\int_{\frac{-\pi}{d}}^{\frac{\pi}{d}}\frac{d^{2} {\bf q} 
dk}{(2\pi)^{3}} \frac{\textstyle
                 |4\pi i d [\hat{z} \times \hat{\bf q}] \cdot 
{\mbox{\boldmath $\rho$}} ({\bf q},k) |^{2}}{
\textstyle q^{2}+\eta_{k}^{2}}+O(\rho^{4})     \label{F1}
\end{eqnarray}
Here we introduced:
\[\theta^{0}({\bf q},k)=-\frac{\textstyle  \eta_{k}^{2} B({\bf q},k) 
}{\textstyle q^{2}+ \eta_{k}^{2}} \]
\[ \eta_{k}^{2}=4\lambda_J^{-2}(\sin^{2}\frac{kd}{2}+\frac{d^{2}}{4 
\lambda_{a b}^{2}}) \]
where the Josephson length is 
\[\lambda_J=\frac{\Phi_{0} d}{4\lambda_{a b} \sqrt{E_J d \pi^{3}}}\]
Since $d\ll \lambda_{a b}$ typically  $\eta_{k}\approx 2/ \lambda_J$ for 
most  $k$ averages below.

The last term in Eq.\ (\ref{F1}) contributes to the longitudinal  
$c_{44}^{l}$ and
 transverse  $c_{44}^{tr}$ part of the tilt moduli. Rewriting the 
 integrand in the form 
\begin{equation}
\int^{\infty}d^{2}{\bf q}
g({\bf q},k)=\sum_{{\bf Q}}
\int^{Q_{0}}d^{2}{\bf q} g({\bf q}+{\bf Q},k)\, ,  \label{sum0}
\end{equation}
using the symmetry of the
hexagonal lattice Eqs.\ (\ref{hex}) as well as 
Eq.\ (\ref{cont}) with 
an upper cutoff $1/\xi$, the tilt moduli  [including the magnetic 
contribution, Eqs.\ (\ref{c440l},\ref{c440t})] can be written as
\begin{mathletters}
\begin{eqnarray}
c_{44}^{l}(q,k)    &=&c_{44}^{l,0}-\frac{2 B\Phi_{0}}{(8\pi 
\lambda_{c})^{2}}\ln\xi^{2}[Q_{0}^{2}+(1+\lambda_{a 
b}^{2}k_{z}^{2})/\lambda_{c}^{2}]\\ 
c_{44}^{tr}(q,k)   &=&c_{44}^{tr,0}+
          \frac{B^{2}}{4\pi} \frac{1}{1+\lambda_{c}^{2} 
q^{2}+\lambda_{a b}^{2}k_{z}^{2}}- 
                  \frac{2 B\Phi_{0}}{(8\pi 
\lambda_{c})^{2}}\ln\xi^{2}[Q_{0}^{2}+(1+\lambda_{a 
b}^{2}k_{z}^{2})/\lambda_{c}^{2}]      \label{c44t}
\end{eqnarray}
\end{mathletters}
where $\lambda_{c}=\lambda_{a b}\lambda_J /d$.
The result for $c_{44}^{tr}(q,k)$ was obtained by GK \cite{GK}.

It is interesting to note  that for $q\rightarrow 0$ and finite 
Josephson coupling $E_J$
the tilt moduli are equal. Note also that the limits $q\rightarrow 
0$ and $E_J \rightarrow 0$ do not commute in the second
term of Eq.\ (\ref{c44t}). In fact we show below that the expansion of Josephson 
term breaks down when 
$E_J \rightarrow 0$.

In summary, the "naive" expansion needs a revision to correct two 
aspects: (i) short scale behavior - expansion is not
allowed in the "$\rho$ circle", and (ii) long scale behavior - generation 
of an $E_J$ independent term when ${\bf q} \rightarrow 0$.

\subsection{Josephson Coupling: Proper Expansion }

We proceed now to a method which avoids the "naive" expansion by 
expanding the cosine term in Eq.\ (\ref{FJos})
 directly in terms of $\rho$.

 We  define $\theta^{n}({\bf r})=\theta^{n,1}({\bf r})+\epsilon^{n}({\bf r})$ and 
expand the Josephson coupling term in Eq.\ (\ref{FJos}) with respect to 
$\epsilon^{n}({\bf r})$ to find an optimal $ \theta^{n,1}({\bf r})$ for which 
the expansion  is allowed:
\begin{eqnarray}
{\cal F}&=&{\cal F}_{v}+\frac{1}{2}\int\frac{d^{2} {\bf q} 
dk}{(2\pi)^{3}} G_{f}^{-1}({\bf q},k) \left[|\theta^{1}({\bf 
q},k)|^{2}+|\epsilon({\bf q},k)|^{2}+\theta^{1}({\bf 
q},k)\epsilon^{*}({\bf q},k)+c.c \right]\nonumber\\
             &-&E_J  \sum_{n}\int d^{2}{\bf r} 
\left[\cos[\theta^{n,1}({\bf r})+\sum_{l}\psi_{l}^{n}({\bf 
r})]-1-\frac{1}{2}\epsilon^{n}({\bf r})^{2}-
\epsilon^{n}({\bf r})
\sin[\theta^{n,1}({\bf r})+\sum_{l}\psi_{l}^{n}({\bf r})]\right]\nonumber\\ 
 & &+O(\epsilon^{4},\epsilon^{2} \rho^{2})      \label{F2}
\end{eqnarray} 
We show below that terms of order $
\epsilon^{2}(\cos[\theta^{n,1}({\bf r})+\sum\psi_{l}^{n}({\bf r})]-1)$
contribute $O(\epsilon^{2} \rho^{2})$ correction to the free energy 
after integration
over ${\bf r}$.

The expansion is most efficient when the term linear in 
$\epsilon({\bf q},k)$ vanishes. This determines $\theta^{1}({\bf 
q},k)$ to be the solution of:
\begin{equation}
\theta^{1}({\bf q},k)=-\frac{\eta_{k}^{2}}{q^{2}}d\sum_{n}\int 
d^{2}{\bf r} 
\sin\left(\theta^{n,1}({\bf r})+\sum_{l}\psi_{l}^{n}({\bf r})\right) 
e^{i{\bf q} {\bf r}+i k n d}.  \label{theta1eq}
\end{equation}
To solve Eq.\ (\ref{theta1eq}) we introduce the functions
 \begin{eqnarray}
     D_{l}^{n}({\bf r})&=&e^{i[\theta^{n,1}_{l}({\bf r})+\psi_{l}^{n}({\bf 
r})]}-1                                       \label{Dln}\\
\delta^{n}_{l}({\bf r})&=&\theta^{n,1}_{l}({\bf r})+C_{l}^{n}
({\bf r}) \label{dln}
\end{eqnarray}
where $C_{l}^{n}({\bf r})$ is defined as
\begin{eqnarray}
C_{l}^{n}({\bf r}) &=&\frac{2 [{\mbox{\boldmath $\rho$}}_{l}^{n}
\times {\bf v}_{l}^{n}({\bf r}) ]_{z}} 
{v_{l}^{n}({\bf r})^{2}+(\rho_{l}^{n})^2 } \nonumber\\                   
    C_{l}({\bf q},k)&=&d\sum_{n} \int d^{2}{\bf r} e^{i{\bf q} 
{\bf r}+iknd}C_{l}^{n}({\bf r})=\nonumber\\
                         &=&4\pi i d\sum_{n} 
\frac{[{\mbox{\boldmath $\rho$}}_{l}^{n} \times \hat{q}]_{z}}
{q}\left(q \rho_{l}^{n} K_{1}(q \rho_{l}^{n})\right) e^{i{\bf q}{\bf 
R}_{l}^{n}+i k n d}
\end{eqnarray}
In the rest of this section we identify expansion parameters 
( Eq.\ (\ref{parameters}) below) which allow a solution of 
Eq.\ (\ref{theta1eq}), and derive the 
free energy expansion (Eq.\ (\ref{F3}) below).

Since the function $C_{l}^{n}({\bf r})$ is close to
 \begin{equation}
  \sin\psi_{l}^{n}({\bf r})  =
           \frac{\textstyle 2[{\mbox{\boldmath $\rho$}}_{l}^{n}
            \times{\bf v_{l}^{n}({\bf r})}]_{z}}{\textstyle
[(v_{l}^{n}({\bf r})^{2}+|{\mbox{\boldmath $\rho$}}_{l}^{n}|^{2})^{2}
-4({\bf v}_{l}^{n}({\bf r})\cdot 
{\mbox{\boldmath $\rho$}}_{l}^{n})^{2}]^{\frac{1}{2}}}
\end{equation}
 for both $v_{l}^{n}\ll \rho_{l}^{n}$ and $v_{l}^{n}\gg 
\rho_{l}^{n}$, the difference between imaginary part of 
$D_{l}^{n}({\bf r})$,
 $ImD_{l}^{n}({\bf r})$, and $\delta^{n}_{l}({\bf r})$ is only on the
"$\rho$-circle", so that
\begin{mathletters}
\begin{eqnarray}
& & \int d^{2}{\bf r}|ImD_{l}^{n}({\bf r})-\delta^{n}_{l}({\bf r})|\sim 
O(\rho^{2})                                     \label{circle1}\\
& &\int d^{2}{\bf r} \sum_{l\neq l'}ImD_{l}^{n}({\bf r})ImD_{l'}^{n}
({\bf r})= \int d^{2}{\bf r} \sum_{l\neq 
l'}\delta^{n}_{l}({\bf r})\delta^{n}_{l'}({\bf r})+O(\rho^{3}/a) . 
\label{circle2}
\end{eqnarray}
\end{mathletters}

We show now that an expansion in $\rho_{l}^{n}$ is possible if the 
following expansion parameters are small
\begin{mathletters}
\label{parameters}
\begin{eqnarray}
\chi &=& \frac{2 d}{\pi 
a^{2}}\int_{1/\lambda_{J}}^{1/a} \frac{d^{2}q}{q^{2}}
\int_{-\pi/d}^{\pi/d}d k
\langle |\rho^{tr}({\bf q},k)|^2 \rangle \ll 1  \hspace{18mm} 
{\mbox if} \:\:\lambda_J\gg a                              \label{chi}\\
\langle \epsilon \rangle                                        
&\ll & 1 \hspace{85mm}   {\mbox if}\:\:\lambda_J\ll a \label{eps}
\end{eqnarray}
\end{mathletters}
where $\langle \rho^{2}\rangle$ is an average of $\rho^{2}$ which is 
diagonal in  ${\bf q},k$.
The case of thermal average is evaluated in section IV.
The parameter $\chi$ controls the expansion of the sine term in 
Eq.\ (\ref{theta1eq}) and is evaluated in Appendix B while 
$\langle \epsilon \rangle = O(\rho ^2)$ results from the
 solution of Eq.\ (\ref{theta1eq}) which is to leading order in 
 $\rho$, so that the term linear in $\epsilon$ in Eq.\ (\ref{F2}) 
 survives and leads to higher order corrections.
  
We claim then that the solution 
$\theta^{n,1}({\bf r})=\sum_{l}\theta^{n,1}_{l}({\bf r})$ of 
Eq.\ (\ref{theta1eq}) (compare with
$\theta^{n,0}({\bf r})$ from "naive expansion") is
\begin{equation}
\theta^{1}({\bf q},k)=\sum_{l}\theta^{1}_{l}({\bf 
q},k)=-\frac{\textstyle \eta_{k}^{2} \sum_{l}C_{l}({\bf 
q},k)}{\textstyle q^{2}+ \eta_{k}^{2}}      \label{theta1}
\end{equation}
so that with Eq.\ (\ref{dln})
\begin{equation}
\delta_{l}^{n}({\bf r})=\frac{d}{\pi}\sum_{m}\int_{-\pi/d}^{\pi/d} d k 
\int_{0}^{\infty}d q\frac{q^{3}\rho_{l}^{m}K_{1}(q 
\rho^{m}_{l})J_{1}(q v_{l}^{m})}{q^{2}+\eta_{k}^{2}}
[\hat{\bf v}_{l}^{n}\times{\mbox{\boldmath $\rho$}}_{l}^{m}]_{z} e^{i 
k(n-m)d}                                        \label{delta}
\end{equation}
where $K_{1}$, $J_{1}$ are conventional Bessel functions.
The function $\delta_{l}^{n}({\bf r})$, in terms of ${\bf 
v}_{l}^{n}={\bf r}-{\bf R}_{l}^{n}$, decays slowly as $1/ v_{l}^{n}$ for 
$a<v_{l}^{n}<\lambda_{eff}$ where $\lambda_J<\lambda_{eff}<\lambda_{c}$ 
depends on the configuration of ${\mbox{\boldmath $\rho$}}_{l}^{n}$, but for 
$v_{l}^{n}>\lambda_{eff}$ it decays as $\exp(-v_{l}^{n} 
/\lambda_{eff})$. The
 exponential decay allows the convergence of the $l$ summations inside 
the sine in Eq.\ (\ref{theta1eq});
however, since the exponential decay sets in at the 
scale $\lambda_{eff}$ which diverges when $E_J\rightarrow 0$,
we expect the expansion parameter $\chi$ of Eq.\ (\ref{chi}) to 
diverge, i.e. the expansion is invalid when $E_J\rightarrow 0$.

The convergence of $\sum_{l}\delta_{l}^{n}({\bf r})$ implies that 
$\sin(\sum_{l} \delta_{l}^{n}({\bf r}))$ can be expanded. More
 precisely, as shown in Appendix B, the condition of Eq.\ (\ref{chi}) 
 leads to Eq. (B3), which together with Eq.\ (\ref{circle1}) yields
\begin{eqnarray}
& &d\sum_{n}\int d^{2}{\bf r} 
\sin\left(\theta^{n,1}({\bf r})+\sum_{l}\psi_{l}^{n}({\bf r})\right) 
e^{i{\bf q} {\bf r}+i k n d}=\nonumber\\
& &d\sum_{n}\sum_{l}\int d^{2}{\bf r} 
Im D^{n}_{l}({\bf r}) e^{i{\bf q} {\bf r}+i k n d}[1+O(\chi)]=\nonumber\\
& &d\sum_{n}\sum_{l}\int d^{2}{\bf r} \delta^{n}_{l}({\bf r})
e^{i{\bf q} {\bf r}+i k n d}[1+O(\chi)]+O(\rho^{2})=\nonumber\\
& &[\theta^{1}({\bf q},k)+\sum_{l}C_{l}({\bf q},k)]
[1+O(\chi)]+O(\rho^{2})                          \label{sum1}
\end{eqnarray}

Substituting in Eq.\ (\ref{theta1eq}) shows that Eq.\ (\ref{theta1})
 is indeed the solution for Eq.\ (\ref{theta1eq}), i.e. it is
the optimal $\theta_{l}^{n,1}$. Furthermore we have
from Eq. (B4), using Eq.\ (\ref{circle2})
\begin{eqnarray}
&\int &d^{2}{\bf r} [\cos(\theta^{n,1}({\bf r})+\sum_{l}\psi_{l}^{n}({\bf 
r}))-1]=\nonumber\\
&\int &d^{2}{\bf r} \left(\sum_{l} Re 
D^{n}_{l}({\bf r})-\frac{1}{2}\sum_{l\neq l'}ImD_{l}^{n}({\bf r}) 
ImD_{l'}^{n}({\bf r})\right)
[1+O(\chi)] =\nonumber\\
&\int &d^{2}{\bf r} \left(\sum_{l} Re 
D^{n}_{l}({\bf r})-\frac{1}{2}\sum_{l\neq l'}\delta_{l}^{n}({\bf r}) 
\delta_{l'}^{n}({\bf r})\right)[1+O(\chi)]
\end{eqnarray}

where $Re D^{n}_{l}({\bf r})$ is the real part of $D^{n}_{l}({\bf r})$. 
Substituting in Eq.\ (\ref{F2}) we obtain
\begin{eqnarray}
{\cal F}&=&{\cal F}_{v}+\left(\frac{1}{2}\int\frac{d^{2} {\bf q} 
dk}{(2\pi)^{3}}(G_{f}^{-1}({\bf q},k)+E_J/d)
\left|\epsilon({\bf q},k) \right|^{2}+\right.  \nonumber  \\
     &+&\frac{E_J}{2d} 
                 \int\frac{d^{2} {\bf q} dk}{(2\pi)^{3}} 
\frac{\textstyle q^{2}
                 |C({\bf q},k) |^{2}}{\textstyle 
q^{2}+\eta_{k}^{2}}-\nonumber\\
        &-&\left.\frac{E_J}{2} \sum_{l,n}\int d^{2}{\bf r}
\left[cos(\theta^{n,1}({\bf r})+ \psi_{l}^{n}({\bf r}) )
+\frac{1}{2}(\delta^{n}_{l}({\bf r}))^{2}-1 
\right]\right)[1+O(\epsilon,\epsilon^{2},\chi)]   \label{F3}
\end{eqnarray}
where $C({\bf q},k)=\sum_{l}C_{l}({\bf q},k) $. 

The balance between the first $|\epsilon({\bf q},k)|^{2}$ term in 
Eq.\ (\ref{F3}) and the $O(\epsilon)$ term leads to $\langle \epsilon 
\rangle \sim \langle \rho ^2 \rangle$.
 The $O(\epsilon^{2})$ term depends on the distribution of
$\epsilon({\bf q},k)$; for thermal average it has a comparable value 
(section IV).

We have identified two types of expansion parameters. The first one, 
$\chi$, is related to the convergence of the $l$
summation of singular vortex phases while the second 
one, $\epsilon$,
 is related to the response of the nonsingular phase.
 For weak Josephson coupling $\lambda_J\gg a$ we find $\chi\gg \langle 
\epsilon \rangle$ so that the expansion parameter is $\chi$ 
while for $\lambda_J\ll a$ we find $\chi\ll \langle \epsilon \rangle$ so 
that the expansion parameter is $\langle \epsilon \rangle$.

 Consider now the function $\delta^{n}({\bf r}) =\sum_{l}\delta_{l}^{n}({\bf r})$.
Since $q\lesssim 1/\rho_{l}^{m}$ (due to $K_{1}$ function)
for $\rho_{l}^{m}\ll \lambda_J$  the dominant integral range with 
$q\approx 1/\rho$  has $q^{2}+\eta_{k}^{2}\simeq q^{2}$. Hence 
$\delta^{n}({\bf r})- C^{n}({\bf r})=O([\rho/\lambda_J]^{2})$  and the last term in
Eq.\ (\ref{F3}) for $ \rho\ll \lambda_J$ can be replaced by
\begin{equation}
\int d^{2}{\bf r}(\cos\psi_{l}^{n}({\bf r}) -1+\frac{1}{2} 
C_{l}^{n}({\bf r})^{2})=-\pi \ln[4e] 
(\rho_{l}^{n})^{2}
\end{equation}
It is straightforward to see that the integral is convergent and 
therefore must be proportional to $(\rho_{l}^{n})^{2}$; the 
coefficient can be found after some algebra.
 
The contribution to the second term in Eq.\ (\ref{F3}) from  
different flux lines can be written in the form

\begin{figure}[htb]
\begin{center}
\includegraphics[bb=72 230 533 600, scale=0.7]{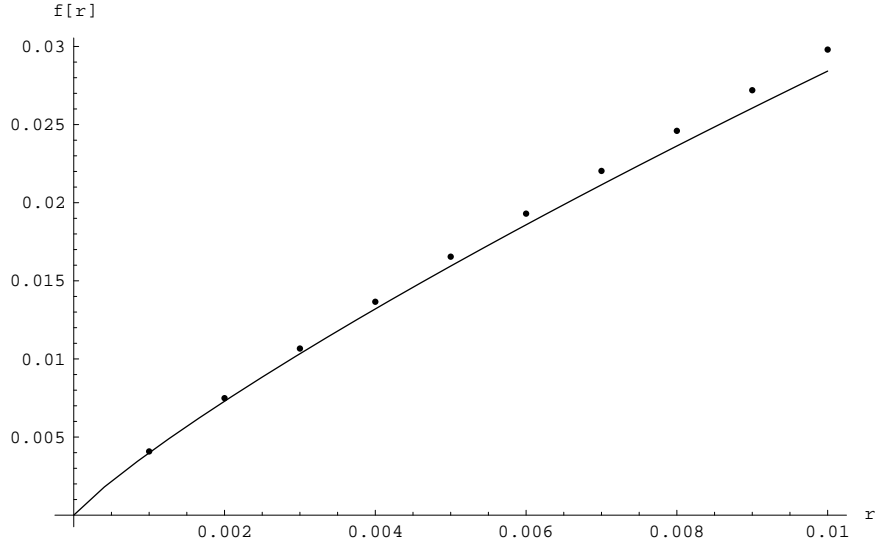}
\end{center}
\caption{Contribution of the Josephson coupling to
the energy of one displaced pancake vortex in one vortex line 
in units of $2\pi E_J \lambda_J^{2}$ ($f=F_1/(2\pi E_J\lambda _J ^2$, 
see appendix C).
Numerical results (dots) and the analytic form Eq. (C2) (line) are shown
  for $r=2\rho^{2}/\lambda_J^{2} \ll 1$.}
\end{figure}

\vspace{10mm}
\begin{eqnarray}
&  &\sum_{l \neq l'} 
\int^{\infty}\int_{\frac{-\pi}{d}}^{\frac{\pi}{d}}\frac{d^{2} {\bf q} 
dk}{(2\pi)^{3}} \frac{q^{2}
C_{l}({\bf q},k)C_{l}^{*}({\bf q},k)}{\textstyle 
q^{2}+\eta_{k}^{2}}\nonumber\\
&=&2 d^{2}\sum_{n,n'}\sum_{l\neq l'}\mbox{\boldmath 
$\rho$}_{l}^{n}\cdot
\mbox{\boldmath $\rho$}_{l}^{n'}\int_{\frac{-\pi}{d}}^{\frac{\pi}{d}} 
d k
K_{0}(\eta_{k}|{\bf R}_{l}^{n}-{\bf R}_{l'}^{n'}|) e^{i 
k(n-n')d}\nonumber\\
&-&2 d^{2}\sum_{n,n'}\sum_{l\neq l'}\rho_{l}^{n} \rho_{l}^{n'}
\cos(\beta_{l}^{n}+\beta_{l'}^{n'})\int_{\frac{-\pi}{d}}^{\frac{\pi}{d}} 
d k K_{2}(\eta_{k}|{\bf R}_{l}^{n}-{\bf R}_{l'}^{n'}|) e^{i 
k(n-n')d}[1+O(\frac{\rho^{2}}{a^{2}})]\label{sum2}
\end{eqnarray}
where $\beta_{l}^{n}$ is the angle between $\mbox{\boldmath 
$\rho$}_{l}^{n}$ and ${\bf R}_{l}^{n}-{\bf R}_{l'}^{n'}$. 
Since $l\neq l'$, the argument of the Bessel functions $K_{0}$, $K_{2}$ 
 is always finite and the limit
 $\rho_{l}^{n}\rightarrow 0$ in this argument can be taken; hence 
$l\neq l'$ terms have a harmonic expansion.

The contribution to the  second term in Eq.\ (\ref{F3}) from  single flux lines, 
i.e. $l=l'$,  can
be computed analytically  for $\rho\ll \lambda_J$
\begin{eqnarray}
& & \int^{\infty}\int_{\frac{-\pi}{d}}^{\frac{\pi}{d}}\frac{d^{2} 
{\bf q} dk}{(2\pi)^{3}} \frac{q^{2}
|C_{l}({\bf q},k)|^{2}}{\textstyle q^{2}+\eta_{k}^{2}}\nonumber\\
&=&2 d^{2}\sum_{n,n'}\mbox{\boldmath $\rho$}_{l}^{n}\cdot
\mbox{\boldmath $\rho$}_{l}^{n'}\int_{\frac{-\pi}{d}}^{\frac{\pi}{d}} 
d k \int_{0}^{\infty} d x \frac{x^{3}
K_{1}(x\sqrt{\frac{\rho_{l}^{n}}{\rho_{l}^{n'}}})
K_{1}(x\sqrt{\frac{\rho_{l}^{n'}}{\rho_{l}^{n}}})}{x^{2}+\eta_{k}^{2}
\rho_{l}^{n}\rho_{l}^{n'}}e^{i k (n-n')d} \nonumber\\
&=&4 \pi d\sum_{n}( \rho_{l}^{n})^{2} \ln(\frac{\lambda_J}{\rho_{l}^{n}})
+4\pi d\sum_{n\neq n'}\mbox{\boldmath $\rho$}_{l}^{n}\cdot 
\mbox{\boldmath $\rho$}_{l}^{n'}e^{-|n-n'|\lambda_J/
\lambda_{c}}/|n-n'| +O(\rho^{4}/\lambda_J^{2})\, .    \label{sum3}
\end{eqnarray}
The last line is obtained by introducing  
${x=q\sqrt{\rho_{l}^{n}\rho_{l}^{n'}}}$ and writing the integral 
 (\ref{sum3}) as
\[\int_{\frac{-\pi}{d}}^{\frac{\pi}{d}} d k\int_{0}^{1}d 
x\frac{x}{x^{2}+\eta_{k}^{2}
\rho_{l}^{n}\rho_{l}^{n'}}e^{i k (n-n')d}
+\int_{0}^{1}\frac{x^{3}\left[
K_{1}(x\sqrt{\frac{\rho_{l}^{n}}{\rho_{l}^{n'}}})
K_{1}(x\sqrt{\frac{\rho_{l}^{n'}}{\rho_{l}^{n}}})-1/x^{2}\right]}{
x^{2}+\eta_{k}^{2}\rho_{l}^{n}\rho_{l}^{n'}}e^{i k (n-n')d}\] 
\[+\int_{1}^{\infty} d x \frac{x^{3}      
K_{1}(x\sqrt{\frac{\rho_{l}^{n}}{\rho_{l}^{n'}}})
K_{1}(x\sqrt{\frac{\rho_{l}^{n'}}{\rho_{l}^{n}}})}{x^{2}+\eta_{k}^{2} 
\rho_{l}^{n}\rho_{l}^{n'}}e^{i k (n-n')d}  \, .\]
In the last two terms  one can put all $\rho_{l}^{n}\rightarrow 0 $ 
since both integrals converge. After separating the $n=n'$ and $n\neq n'$ 
terms and integrating over 
$k$ the result Eq.\ (\ref{sum3}) is obtained. 
 In Appendix 
C we consider displacement of a single pancake vortex in one flux line 
and demonstrate the agreement between a numerical exact evaluation of 
Eq.\ (\ref{F3}) and the analytic expansion (see Fig. 2). We also show
in Appendix C that  for the single pancake displacement an 
expansion in $\rho$ is possible also for $\lambda_J <\rho <a$. 

We have shown in Eq.\ (\ref{sum3}) that in general there are
$(\rho_{l}^{n})^{2}\ln\rho_{l}^{n}$ terms in
the energy expansion, confirming the anticipation by GK \cite{GK}. 
Thus, strictly speaking the elastic constants are 
ill defined. However, $\ln\rho_{l}^{n}$ is a slowly varying function so
that replacing it by an average value $\ln\bar{\rho}$ should yield the 
main nonlinear correction. At finite temperatures 
$\bar{\rho}=\langle \rho^{2}\rangle ^{1/2}$
would be a thermal average. This procedure is tested for the single pancake 
displacement (Appendix  C) and is found to be in a good agreement 
with the exact thermal average.

\subsection{Effective Elastic Constants}

Effective elastic constants are obtained by replacing non-harmonic 
terms in $\rho_{l}^{n}$ by an average value 
$\bar{\rho}$, which is to be determined self consistently  e.g by a 
thermal average   $\bar{\rho}=\langle \rho^{2}\rangle ^{1/2}$.
We introduce the 
effective
 singular phase difference $\psi_{l}^{n}({\bf r})^{eff}$ which leads 
to effective  $C_{l}^{n,eff}({\bf r})$ functions
\begin{eqnarray}
         \sin\psi_{l}^{n}({\bf r})^{eff}&=&\frac{\textstyle 2[
         {\mbox{\boldmath $\rho$}}_{l}^{n} 
\times{\bf v_{l}^{n}({\bf r})}]_{z}}{\textstyle [(v_{l}^{n}({\bf 
r})^{2}+ \bar{\rho}^{2})^{2}-4({\bf v}_{l}^{n}({\bf r})\cdot 
\hat{\rho}_{l}^{n})^{2} \bar{\rho}^{2}]^{\frac{1}{2}}}\nonumber\\
C_{l}^{n,eff}({\bf r})&=&\frac{2 [{\bf v}_{l}^{n}({\bf r}) \times 
{\mbox{\boldmath $\rho$}}_{l}^{n}]_{z}} 
{v_{l}^{n}({\bf r})^{2}+ \bar{\rho}^{2}}\nonumber\\
       C^{eff}({\bf q},k)&=&\frac{4\pi i d[\hat{z} \times \hat{\bf 
q}] \cdot \mbox{\boldmath $\rho$} ({\bf q},k) }{q}  \bar{\rho}q 
K_{1}(\bar{\rho}q) 
\end{eqnarray}
This approximation simplifies significantly all computations of 
averages because the energy can  be written in the harmonic form
\begin{eqnarray}
{\cal F}&=&{\cal F}_{v}+\frac{1}{2}\int\frac{d^{2} {\bf q} 
dk}{(2\pi)^{3}}(G_{f}^{-1}({\bf q},k)+E_J/d)
\left|\theta({\bf q},k)-\theta^{1,eff}({\bf q},k) \right|^{2} 
\nonumber  \\
             &+&\frac{1}{2} (4\pi)^{2} E_J d                 
\int^{\infty}\int_{\frac{-\pi}{d}}^{\frac{\pi}{d}}\frac{d^{2} {\bf q} 
dk}{(2\pi)^{3}} 
\frac{\textstyle   | [\hat{z} \times \hat{\bf q}] \cdot 
\mbox{\boldmath $\rho$} ({\bf q},k) |^{2} q^{2} 
\bar{\rho}^{2}K_{1}^{2}(q \bar{\rho})}
{\textstyle q^{2}+\eta_{k}^{2}} \nonumber \\
             &+&\frac{\pi E_J 
}{d}\int^{Q_{0}}\int_{-\frac{\pi}{d}}^{\frac{\pi}{d}}\frac{d^{2} 
{\bf q} dk}{(2\pi)^{3}}d a^{2}\ln(4 e)|\mbox{\boldmath $\rho$}({\bf 
q},k)|^{2}                            \label{F4}
 \end{eqnarray}
where we define
\[\theta^{1,eff}({\bf q},k)=-\frac{\textstyle \eta_{k}^{2} 
C^{eff}({\bf q},k)}{\textstyle
q^{2}+ \eta_{k}^{2}}  \, . \]   
 
To derive Eq.\ (\ref{F4}) the "Proper Expansion" is used with the new 
$\sin\psi_{l}^{n}({\bf r})^{eff}$ . The result is similar to 
replacing $\rho$ by $\bar{\rho}$ into the  coefficients of 
non-harmonic terms in Eq.\ (\ref{F3}).

Using Eqs.\ (\ref{cont},\ref{sum0})  the second term in 
Eq.\ (\ref{F4}) can be written as
\begin{eqnarray}
& &\int^{\infty}\int_{\frac{-\pi}{d}}^{\frac{\pi}{d}}\frac{d^{2} {\bf 
q} dk}{(2\pi)^{3}} 
\frac{\textstyle   | [\hat{z} \times \hat{\bf q}] \cdot 
\mbox{\boldmath $\rho$} ({\bf q},k) |^{2} q^{2} 
\bar{\rho}^{2}K_{1}^{2}(q\bar{\rho})}
{\textstyle q^{2}+\eta_{k}^{2}}= \nonumber\\
&=& \int^{Q_{0}}\int_{\frac{\pi}{d}}^{\frac{\pi}{d}} \frac{d^{2} {\bf 
q} dk}{(2\pi)^{3}}  \left[
\frac{\textstyle |[\hat{z} \times \hat{\bf q}] \cdot \mbox{\boldmath 
$\rho$} ({\bf q},k) |^{2} q^{2} \bar{\rho}^{2}K_{1}^{2}(q\bar{\rho})}
{\textstyle q^{2}+\eta_{k}^{2}}\right.+\nonumber \\
&+&\left. |\mbox{\boldmath $\rho$} ({\bf q},k) 
|^{2}\frac{1}{2}\frac{1}{Q_{0}^{2}}\int _ {Q_{0}\bar{\rho}} 
^{\infty}\frac{x^{3}K_{1}^{2}(x) d x} {x^{2}+\eta_{k}^{2} 
\bar{\rho}^{2}}\right] 
\end{eqnarray}
where  the last integral has the analytic fit
\[\int _ {Q_{0}\bar{\rho}} ^{\infty}\frac{x^{3}K_{1}^{2}(x) d x} 
{x^{2}+\eta_{k}^{2} \bar{\rho}^{2}}=-\frac{1}{2} \ln \left( \frac{ 
\bar{\rho}^{2}\eta_{k}^{2}(1+Q_{0}^{2}\eta_{k}^{-2})} 
{ \bar{\rho}^{2}\eta_{k}^{2}+1} \right)\]
The effective free energy  of the vortex lattice can now be written in
 the harmonic form with
effective transverse, $c_{44}^{tr}({\bf q},k)$, and longitudinal, 
$c_{44}^{l}({\bf q},k)$,
 tilt moduli 
\begin{mathletters}
\label{c44}
\begin{eqnarray}
c_{44}^{l}(q,k)  &=&c_{44}^{0,l}(q,k) -
          \frac{2 B\Phi_{0}}{(8\pi \lambda_{c})^{2}}
\ln\left[ (\bar{\rho}^{2}/4 e)
( Q_{0}^{2}+(1+\lambda_{a b}^{2} k_{z}^{2})/\lambda_{c}^{2})
\right]                                        \label{c44le}\\
c_{44}^{tr}(q,k) &=&c_{44}^{0,tr}(q,k) +
               \frac{B^{2}}{4\pi} \frac{1}{1+\lambda_{c}^{2} 
q^{2}+\lambda_{a b}^{2}k_{z}^{2}}\\
&-&\frac{2 B\Phi_{0}}{(8\pi \lambda_{c})^{2}}
\ln\left[ (\bar{\rho}^{2}/4 e)
( Q_{0}^{2}+(1+\lambda_{a b}^{2} k_{z}^{2})/\lambda_{c}^{2})
\right]                                         \label{c44te}
\nonumber
\end{eqnarray}
\end{mathletters}
It is seen that  $c_{44}(q,k)$ of the
 ''naive'' expansion  is now corrected by replacing $\xi^{2}$ with
 $\bar{\rho}^2/4 e$. We have assumed implicitly that ${\bar 
\rho}>\xi$, otherwise the "$\rho$ circle" is within the
vortex core area where $J$ is reduced, i.e. the starting model 
Eq.\ (\ref{LD}) should be modified.

\section{Thermal Averages}

To determine the effective tilt moduli Eqs.\ (\ref{c44}),
 as well as the 
conditions for the expansion Eqs.\ (\ref{parameters}) we need to
evaluate the thermal average of the relative displacement of pancake 
vortices
\begin{equation}
\bar{\rho}^{2}=T\int^{Q_{0}}\int^{\frac{\pi}{d}}\frac{d^{2}{\bf q} 
d k}{(2\pi)^{3}}            
\left[\frac{\sin^{2}(\frac{kd}{2})}{q^{2}c_{11}+k_{z}^{2} c_{44}^{l}}
 +\frac{\sin^{2}(\frac{kd}{2})}{ q^{2}c_{66}+k_{z}^{2} 
c_{44}^{tr}}\right] 
\end{equation}
For weak magnetic fields where $\lambda_J\ll a$ the Josephson 
contribution to $c_{44}^{l}$ and $c_{44}^{tr}$ is dominant and the 
latter dominate the integrals, leading to
\begin{equation}
 \bar{\rho}^{2} \approx   4 \frac{T}{\tau }\lambda_J^{2}/
\ln(\frac{e\tau \ln(4e)}{T})[1+O(\frac{\lambda_J^{2}}{a^{2}},
\frac{a^{2}}{\lambda_{a b}^{2}})]                    \label{rhoav}
 \end{equation}
where $\tau=\Phi_{0}^{2}d/(4\pi^{2} \lambda_{ab}^{2})$ and  
$T\ll\tau$ is  assumed. 
The expansion parameter for $\lambda_J\ll a$ (Eq.\ (\ref{eps})) together with 
Eq.\ (\ref{rhoav}) yields, in fact, the expansion condition $T\ll\tau$.
We also find by numerical integration that the displacement average is
 $\bar{u^{2}}\approx  \bar{\rho}^{2}$. 
 
 We note that $\bar{\rho}^{2}$ is nonlinear in $T$ due to the $\ln T$ 
 factor in Eq.\ (\ref{rhoav}). Thus data on $\bar{\rho}^{2}$, e.g. by a 
 Debye-Waller term in neutron scattering may probe the $\ln T$ 
 factor in weak fields, i.e. $\lambda_J\ll a$.
 
For strong magnetic fields where $\lambda_{a b}\gg\lambda_J\gg a$
the Josephson contribution to the last terms in Eq.\ (\ref{c44})
can be ignored, i.e. $c_{44}^{tr}\approx c_{44}^{l}\approx 
c_{44}^{0,l}$, leading to
\begin{equation}
\bar{\rho}^{2} \approx \frac{T}{\tau}a^{2}\ln\frac{9\pi\lambda_J^{2}}{32 
a^{2}}[1+O(\frac{a^{2}}{\lambda_J^{2}},\frac{\lambda_J^{2}}{\lambda_{a b}^{2}})]
\end{equation}
where $k=\pi/d$ dominates the integral; here also $\bar{u^{2}}
\approx  \bar{\rho}^{2}$. In this case the thermal average of 
$ \bar{\rho}^{2}$ has the usual linear temperature dependence

We proceed to evaluate the expansion parameters Eqs.\ (\ref{parameters}) which 
determine the validity range of our expansion. For $\lambda_J\gg a $ we 
need $\chi$ of Eq.\ (\ref{chi}),
\begin{equation}
\chi= T\frac{2\pi^{2}d^{2}}{a^{4}}
\int_{1/\lambda_J}^{1/a}\int_{-\pi/d}^{\pi/d}\frac{d^{2}{\bf q} d
k}{(2\pi)^{3}} \frac{k_{z}^{2}}{q^{2}( c_{66}q^{2}+c_{44}^{tr}
({\bf q},k)k_{z}^{2})}
\end{equation}
while for $\lambda_J\ll a $ we evaluate $\langle \epsilon \rangle$ directly with 
Eq.\ (\ref{rhoav}).
We note that the thermal average yields
$\langle \epsilon^{2}\rangle=(2 T/\tau)\ln\lambda_J/\xi$ so that
$\langle \epsilon^{2}\rangle \approx \langle \epsilon \rangle$.

 We find then
that the effective harmonic expansion is valid at temperatures below 
$T^{b}$,
\begin {mathletters}
\begin{eqnarray}
T^{b} & \approx      &\case{1}{2}T_{d}/\ln (2\lambda_J /\lambda_{a b})
	\hspace{33mm}{\mbox if}   \hspace{5mm} a,\lambda_{a b}\ll \lambda_J 
	\label{Tb1}\\
T^{b}&\approx      &\tau a /(\pi \lambda_J )
\hspace{47mm} {\mbox if } \hspace{5mm}   a \ll\lambda_J\lesssim \lambda_{a b} 
\label{Tb2}\\
T^{b}&\approx      &\case{1}{2}\tau/\ln (\lambda_J /\xi ) 
\hspace{41mm}{\mbox if } \hspace{5mm}   \lambda_J \ll a
\end{eqnarray}
\end {mathletters}
Here we defined \cite{Daemen,Baruch} the decoupling 
temperature $T_{d}=\tau a^{2}\ln(a/d)/(4\pi \lambda_{ab}^{2})$ for 
the range $a,\lambda_{a b}\ll \lambda_J$. 
We note that the form of $\chi$ Eq.\ (\ref{chi}) involves precisely 
the fluctuations that lead to the decoupling transition
\cite{Daemen,Baruch}. It is 
therefore expected that $T^b$ is related to $T_d$ for the case when 
$\chi$ is the relevant expansion parameter, i.e. $a \ll \lambda_J$. 
For both cases of Eqs.\ (\ref{Tb1},\ref{Tb2}) the decoupling 
temperature is indeed close to $T^b$ (even for
$a \ll\lambda_J\lesssim \lambda_{a b}$ where the decoupling transition 
becomes first order \cite{Daemen}). For $a \gg \lambda_J$ we expect 
that fluctuations of the non-singular part of the Josephson phase 
($\epsilon({\bf q},k)$) dominate so that the low temperature 
instability involves melting rather than decoupling.

\section{Conclusion}

We present in this work a proper expansion for defining elastic 
constants. Both deficiencies of the "naive" expansion, 
when corrected, lead to interesting physical consequences. The first 
difficulty is that a simple expansion at short scales, 
the "$\rho$ circle" in Fig. 1, is not possible. The proper expansion 
shows a non-harmonic $\rho^{2}\ln \rho$ term so that, 
strictly speaking, $c_{44}$ is ill defined for displacements 
$\rho>\xi$ (for $\rho<\xi$ the Josephson coupling $E_J$ should be 
modified by the reduced order parameter in the vortex core, an effect 
which is neglected in the Lawrence-Doniach model, Eq.\ (\ref{LD})).

We find that effective elastic constants can be defined by replacing 
$\ln\rho$ by $\ln\bar{\rho}$, where $\bar{\rho}$ is
 thermal average $\bar{\rho}=\langle \rho^{2}\rangle ^{1/2}$. This leads to 
replacing $\xi $ of the "naive" expansion by 
$\bar{\rho}/\sqrt{4 e}$ in the effective $c_{44}$, Eqs.\ (\ref{c44}). Since 
$\langle \rho^{2}\rangle \sim T$ this effect can show up as a 
$\ln T$ factor in a direct measurement of $c_{44}$. Furthermore, when 
$\lambda_{J}\ll a$ the Josephson contribution dominates in the tilt 
moduli and $\langle \rho^{2}\rangle \sim T/\ln(\tau /T)$. This 
temperature dependence may be observable via a Debye-Waller factor in 
neutron scattering.

The second deficiency of the "naive" expansion is that a $E_J$ 
independent term is generated from the Josephson term when 
$q\rightarrow 0$. This difficulty relates to the expansion parameter 
$\chi$ of Appendix B - the summation on flux lines
converges only
beyond a scale $\sim \lambda_J$ so that $\chi\sim \rho ^2 \ln \lambda_{J}$.
 When $E_J\rightarrow 0$ the range where the 
$\rho$ expansion is valid, 
$\chi\ll 1$, vanishes as $1/\ln E_J$. Thus at $q\rightarrow 0$ a long 
range effect of many flux lines invalidates 
the $\rho$ expansion. In practice one needs $\lambda_J\gg\lambda_{a b}$ for 
this effect to be noticeable, and the harmonic 
expansion is then limited to $T< \case{1}{2}T_{d}/\ln (2\lambda_J 
/\lambda_{a b})$.

The $q\rightarrow 0$ difficulty is in fact resolved by either a self 
consistent harmonic approximation \cite{Daemen} 
or by a renormalization group method \cite{Baruch}. In both cases the 
cosine 
function is not expanded (although $\psi_{l}^{n}({\bf r})$ is
 expanded as $\psi_{l}^{n}({\bf r}) = -2{\bf \nabla} \alpha({\bf r}-{\bf 
R}_{l}^{n}) \cdot {\mbox{\boldmath $\rho$}}_{l}^{n} $) 
leading to a decoupling temperature $T_{d}$. For $T<T_{d}$ $E_J$ is 
renormalized to a finite value 
$J^{R}\approx T (\xi ^2 E_J/T)^{1/(1-T/T_{d})}$ which can be expanded 
when $T\ll T_{d}/\ln(T/
\xi ^2 E_J)$, equivalent to our expansion parameter. 
For $T>T_{d}$ the renormalized $E_J$ vanishes and the $\rho$
expansion is clearly invalid.

The expansion parameter is related to $T_{d}$ only for $\lambda_J\gg a$, 
while for $\lambda_J\ll a $ the expansion is valid for
 $T<\case{1}{2}\tau/\ln(\lambda_J/\xi)$. 
We expect that in the latter case Josephson fluxons with width $\lambda_J$
 can form loops in between layers and lead to melting of the
 flux lattice. Thus for $\lambda_J\ll a $ the dominant instability is melting 
while for $\lambda_J\gg a$ it is decoupling; in the latter case the
 lattice at $T>T_{d}$
 (held by magnetic coupling) melts at a higher temperature.

In recent experiments on BSCCO \cite{Fuchs} the phase diagram has 
shown a number of low temperature phases 
(related to disorder \cite{Baruch}) while at $T>40 K$, where thermal 
fluctuations dominate, the transition to a vortex
 liquid phase is of two types:
(i) At $B<500 G$ a first order transition with no further transitions 
at highr temperatures, and (ii) at 
$500 G < B < 900 G$ a first order transition 
followed by another transition 
where surface barriers are reduced. For BSCCO $\lambda_J$ is estimated as 
\cite{Kes} $800-2400 \AA$ while at $B=500G$ $a=2000 \AA$. It is then 
consistent to consider the $B<500G$ transition as melting
 ($\lambda_J< a$) while at $B>500 G$ decoupling dominates ($\lambda_J>a$) with 
melting at a higher temperature.

\vspace{10mm}
{\bf Acknowledgments}: 
This research was supported by the Israel Science Foundation founded 
by the Israel Academy of Sciences and Humanities.

\renewcommand{\theequation}{A\arabic{equation}}
\setcounter{equation}{0}

\subsection{Appendix. Tutorial Example}

Let consider two superconducting layers with Josephson coupling 
between
them and  only one pancake vortex on each  
layer. The  Lawerence-Doniach free energy in the simple case of
$e\rightarrow 0$
has the form:
\begin{equation}
 {\cal E}=\int d^{2}{\bf r}(\sum_{n=1,2}[{\bf 
\nabla}\phi_{n}({\bf r})]^{2}-\lambda_J^{-2}
[\cos\left(\phi_{2}({\bf r})-\phi_{1}({\bf r})\right)-1])
\end{equation}

We now decompose the superconducting phase $\phi^{n}({\bf r})$ to the 
nonsingular $\phi^{n}_{0}({\bf r})$ and
singular part:
 \[\phi_{n}({\bf r})=\phi_{n}^{0}({\bf r})+\alpha({\bf r}-{\bf R}_{n})\]
where $\alpha({\bf r})=\arctan(y/x)$ and ${\bf R}_{n}$ is the vortex 
position
on the $n$-th layer.\
Define new $\theta({\bf r})=[\phi_{2}^{0}({\bf r})-\phi_{1}^{0}({\bf r})]$ and
$\psi({\bf r})=\alpha({\bf r}-{\bf R}_{2})-\alpha({\bf r}-{\bf R}_{1})$ to write 
the  
free energy in the form:
\begin{equation}
{\cal E}=\int d^{2}{\bf r}\left(\frac{1}{2}[{\boldmath 
\nabla}\theta({\bf r})]^{2}-\lambda_J^{-2}
[\cos(\theta({\bf r})+\psi({\bf r}))-1]\right)+{\cal E}_{0}
\end{equation}
where ${\cal E}_{0}$ part contains magnetic interaction between 
vortices (we do not consider it here) and
$\frac{1}{2} \int d^{2}{\bf r}[{\boldmath 
\nabla}(\phi_{2}^{0}({\bf r})+\phi_{1}^{0}({\bf r}))]^{2}$ part which can be 
integrated out.

We can perform an expansion of the cosine term with respect to the {\em 
nonsingular} phase $\theta({\bf r})$ since it can be shown 
selfconsistently that
 $|\theta({\bf r})|^{2}\sim \lambda_J^{-2} \rho^{2} \ll 1$ for small relative 
displacements of vortices:
\[{\mbox{\boldmath $\rho$}}=[{\bf R}_{2}-{\bf R}_{1}]/2\]
So we can write:
 \[{\cal E}=\frac{1}{2}\int d^{2}{\bf r}\left([{\boldmath 
\nabla}\theta({\bf r})]^{2}+\lambda_J^{-2} [\theta({\bf r})]^{2}-2\lambda_J^{-2}
[\cos\psi({\bf r})-1]+2\lambda_J^{-2}\theta({\bf r})\sin 
\psi({\bf r})\right)+{\cal 
E}_{0}+O(\theta^{2}\rho^{2},\theta^{4})\]
where
\[\sin\psi({\bf r})=\frac{2[{\mbox{\boldmath $\rho$}}\times {\bf 
v}]_{z}}{[(\rho^{2}+ v^{2})^{2}-4({\bf v}\cdot {\mbox{\boldmath 
$\rho$}})^{2}]^{\frac{1}{2}}}\]
\[{\bf v}={\bf r}-[{\bf R}_{1}+{\bf R}_{2}]/2\]
After shifting $\theta$ by the linear term in $\theta$ term, 
substituting $\sin \psi({\bf r})$ by
$C({\bf r})= 2[{\mbox{\boldmath $\rho$}}\times 
{\bf v}]_{z}/(\rho^{2}+ v^{2})$ we obtain:
\begin{eqnarray}
{\cal E}&=&\frac{1}{2}\int \frac{d^{2}{\bf q}}{(2\pi)^{2}}\left(
(q^{2}+\lambda_J^{-2}) |\theta({\bf q})+\frac{ C({\bf 
q})}{(1+q^{2}\lambda_J^{2})}|^{2} +\frac{ q^{2}|C({\bf 
q})|^{2}}{1+q^{2}\lambda_J^{2}}\right) \nonumber\\
             &-&\lambda_J^{-2}\int d^{2}{\bf r} [ \cos\psi({\bf r})+\frac{1}{2} 
(C({\bf r}))^{2}-1]+O[\theta \rho^{2}]+{\cal E}_{0}
\end{eqnarray}
Where the $\theta[\sin\psi-C]$ term contributes a $O(\theta \rho^{2})$ 
correction to the energy.

It can be calculated analytically that:
\begin{eqnarray}
\int d^{2}{\bf r} [ \cos\psi({\bf r})+\frac{1}{2} (C({\bf r}))^{2}-1] &=& 
-\pi \ln[4 e] \rho^{2}   \\
\frac{1}{2}\int \frac{d^{2}{\bf q}}{(2\pi)^{2}}\frac{q^{2}|C({\bf 
q})|^{2}}{1+q^{2}\lambda_J^{2}} &=&
-\frac{1}{8\pi}\rho^{2}\ln\frac{\lambda_J^{-2}\rho^{2}}
{1+\lambda_J^{-2}\rho^{2}}
+O[\lambda_J^{-4}\rho^{4}]
\end{eqnarray}

This shows the presence of the anharmonic term $\rho^{2}\ln\rho$
 in the energy expansion for two superconducting layers. We obtain 
in section  IIIC this anharmonicity  in the more general
 case of a vortex lattice in a layered  superconductor.

\renewcommand{\theequation}{B\arabic{equation}}
\setcounter{equation}{0}

\subsection{ Appendix. Parameter of Expansion}

Let us introduce a 2-D lattice with ${l}$ the unit cell index and use 
definitions Eq.\ (\ref{vln}) and Eq.\ (\ref{Dln}) for ${\bf v}_{l}^{n}$
and $D_{l}^{n}({\bf r})$.

Consider ${\bf r}$ in  the $l^{*}$ unit cell with ${\bf r}={\bf 
R}_{l^{*}}+{\bf v}_{l^{*}}^{n}$, so that
  $|{\bf v}_{l^{*}}^{n}|<a$. Since
$\delta_{l}^{n}({\bf r})=\bar{\delta}^{n}_{l}(
{\bf v}_{l}^{n} )$ depends
on ${\bf r}$ only through ${\bf v}_{l}^{n}={\bf r}-{\bf R}_{l}$, (see 
Eq.\ (\ref{delta})), we can write
\begin{eqnarray}
\chi^{n}_{I}&=& \sum_{l\neq l^{*}}Im D_{l}^{n}({\bf r})= 
\sum_{l\neq l^{*}}\bar{\delta}^{n}_{l}({\bf v}_{l}^{n})+
\sum_{l\neq l^{*}}(Im D_{l}^{n}({\bf r})-\delta_{l}^{n}({\bf r}))\nonumber\\
                     &=&-\frac{d}{\pi}\sum_{l\neq 
l^{*}}\int_{-\pi/d}^{\pi/d}d k \eta_{k} 
K_{1}(\eta_{k} |{\bf R}_{l}-{\bf R}_{l^{*}}|)
e^{i k n d}\frac{[{\bf v}_{l}^{n}
\times\mbox{\boldmath $\rho$}_{l}(k)]_{z}}
{ |{\bf R}_{l}-{\bf R}_{l^{*}}|}\nonumber\\
& &\times (1+O[\frac{v_{l^{*}}}{ |{\bf 
R}_{l}-{\bf R}_{l^{*}}|}])  +O(\frac{\rho^{2}}{a^{2}})     \\
\chi^{n}_{R}&=&\sum_{l\neq l^{*}}Re D_{l}^{n}=\frac{1}{2}\sum_{l\neq 
l^{*}}   [\bar{\delta}_{l}^{n}({\bf 
v}_{l}^{n})]^{2}+O(\frac{\rho^{2}}{a^{2}})\nonumber\\
                       &=&\frac{d^{2}}{2\pi^{2}}\sum_{l\neq 
l^{*}}\left[\int_{-\pi/d}^{\pi/d}d k \eta_{k} K_{1}(\eta_{k} |{\bf 
R}_{l}-{\bf R}_{l^{*}}|)
\frac{[({\bf R}_{l}-{\bf
R}_{l^{*}})\times\mbox{\boldmath $\rho$}_{l}(k)]_{z}}{|{\bf 
R}_{l}-{\bf
R}_{l^{*}}|}e^{i k n d}\right]^{2}  \nonumber\\
&&\times (1+O[\frac{v_{l^{*}}}{ |{\bf R}_{l}-{\bf R}_{l^{*}}|}])                                    
+O(\frac{\rho^{2}}{a^{2}})
\end{eqnarray}

Using the expansion
\[\prod_{l}(1+D_{l}^{n})=1+\sum_{l} D_{l}^{n} +\frac{1}{2 
!}\sum\sum_{l\neq l'}D_{l}^{n} D_{l'}^{n}+
\frac{1}{3 !}\sum\sum\sum_{l\neq l'\neq l''}D_{l}^{n} D_{l'}^{n} 
D_{l''}^{n}
+ \cdots  \]
we obtain for the right hand side of Eq.\ (\ref{theta1eq})
\begin{eqnarray}
& &\int d^{2}{\bf r} Im \left[\prod_{l}(1+D_{l}^{n}({\bf r}))\right] e^{i 
{\bf q} {\bf r}}
    =\sum_{l^{*}}\int^{a} d^{2}{\bf 
v}_{l^{*}}Im\left[(1+D_{l^{*}}^{n})\prod_{p\neq l^{*}}(1+ 
D_{p}^{n})\right]
e^{i{\bf q} {\bf r}}\nonumber\\
&=&\sum_{l^{*}}\int^{a} d^{2}{\bf v}_{l^{*}}\left[Im 
D_{l^{*}}^{n}(1+O( [\chi^{n}_{I}]^{2},\chi^{n}_{R}))
+(1+Re D_{l^{*}}^{n})(\sum_{l\neq l^{*}}Im D_{l}^{n}+O( 
[\chi^{n}_{I}]^{2},[\chi^{n}_{R}]^{2}))\right]
e^{i{\bf q}{\bf r}}\nonumber\\
&=& \int d^{2}{\bf r}\sum_{l}Im D_{l}^{n} e^{i{\bf q}{\bf r}}
[1+O(\frac{\rho^{2}}{a^{2}}, [\chi^{n}_{I}]^{2},\chi^{n}_{R})]
\end{eqnarray}
here $\int^{a} d^{2}{\bf v}_{l}ReD_{l}=O(\rho^{2})$ is used and  
$\int^{a}$ means integration over the unit cell.

For the Real part of (B2) we obtain 
\begin{equation}
\int d^{2}{\bf r} (Re \prod_{l}(1+D_{l})-1)=\int d^{2}{\bf r}[ 
\sum_{l}Re D_{l}-\frac{1}{2}\sum_{l\neq l'}Im D_{l} Im D_{l'}](
1+O([ \chi^{n}_{I}]^{2},\chi^{n}_{R}))
\end{equation}

Note, that the expansion parameter  depends on the configuration 
$\mbox{\boldmath $\rho$}_{l}^{n}$. We consider averages
of $ [\chi^{n}_{I}]^{2}$ and $\chi^{n}_{R}$ which are diagonal in 
$\rho({\bf q},k)$, e.g. as in thermal average. We define the expansion 
parameter $\chi = \langle [\chi^{n}_{I}]^{2}\rangle$ and obtain the form
\[
\chi = \frac{2 d}{\pi 
a^{2}}\int_{BZ} \frac{d^{2}{\bf q}}{q^{2}}
\int_{-\pi/d}^{\pi/d}d k
\langle |\rho^{tr}({\bf q},k)|^{2}\rangle[J_{0}(a q)-J_{0}(q/\eta_{k})]^{2}
\]
which for $\lambda_J\gg a$ reduces to Eq.\ (\ref{chi}).
For $\lambda_J\gg a$ this is the relevant expansion parameter since 
from Eq.\ (\ref{eps}) $\chi \gg
\langle \epsilon \rangle$; furthermore, the other expansion parameter
\begin{equation}
<\!\!\chi^{n}_{R}\!\!>=2\pi\frac{<\!\![\rho^{n}]^{2}\!\!>}{a^{2}}
\ln\frac{\lambda_J}{a}
\end{equation}
is seen to satisfy $\langle \chi^{n}_{R} \rangle \ll \chi$. 

In the case of $\lambda_J\ll a$ the average yields 
\begin{equation}
\chi \approx \langle \chi^{n}_{R}\rangle \approx 
\bar{\rho}^{2}/4a^2 
\end{equation}
which is much smaller then the other expansion parameter of Eq.\ 
(\ref{eps}).   Thus for 
$\lambda_J\ll a$ the relevant expansion parameter is $\langle \epsilon 
\rangle$.

\renewcommand{\theequation}{C\arabic{equation}}
\setcounter{equation}{0}

\subsection{Appendix. One Displaced Vortex Point  in One Vortex Line}

In the case of one displaced pancake vortex on the layer  $n=0$, in one 
vortex line, $l=0$, so only
$\mbox{\boldmath $\rho$}_{0}^{0}=-\mbox{\boldmath 
$\rho$}_{0}^{1}=\mbox{\boldmath $\rho$}$ are exist, we use Eq.\ 
(\ref{theta1}) in Eq.\ (\ref{F3})
 to evaluate the energy numerically.
The result for the  energy ${\cal F}_{1}$  (without  magnetic part  
${\cal F}_{v})$  is shown by the dots in Fig. 2.

The numerical result can be fitted as:
\begin{equation}
{\cal F}_{1}= 2\pi E_J [1.6\rho^{2}  - 1.04 \rho^{2} 
\ln(\frac{2\rho^{2}/ (\lambda_J^{2})}{2\rho^{2}/ (\lambda_J^{2})+1})]
\end{equation}

For $\rho \ll \lambda_J$ we can write the energy (without
magnetic part)  analytically by using Eq.\ (\ref{sum2})
\begin{equation}
{\cal F}_{1}^{\rho \ll \lambda_J}=2\pi E_J
[\left(\ln\frac{8}{e}\right) \rho^{2} - \rho^{2} \ln
(2\rho^{2}/ (\lambda_J^{2})) ]
\end{equation}

We compare in  Fig. 2  the analytic result (line) with 
numerical calculations (dots).

If we put thermal average of $\rho^{2}$, $\bar{\rho}^{2}$ , into the 
non-harmonic coefficient of $\rho^{2}$,
\begin{equation}
{\cal F}_{1}^{eff}= 2\pi E_J [1.6 \rho^{2}  -1.04 
\rho^{2} \ln(\frac{2\bar{\rho}^{2}/
(\lambda_J^{2})}{2\bar{\rho}^{2}/ (\lambda_J^{2})+1}) ]
\end{equation}
and compute the average of $\rho^{2}$ using either effective ${\cal 
F}_{1}^{eff} $
 or exact ${\cal F}_{1}$ energy we find very similar results.
Indeed, the approximation is good because 
most contribution to the average is from the region of displacements 
near the
average displacement.

\end{document}